\documentclass[twocolumn,showpacs,preprintnumbers,amsmath,amssymb]{revtex4}
\usepackage{graphicx,color}
\usepackage{bm}% 
\newcommand{\eee}{{\bf e}}

\newcommand{\kk}{{\bf k}}

\newcommand{\vv}{{\bf v}}

\newcommand{\rr}{{\bf r}}

\newcommand{\Aa}{{\cal A}}

\newcommand{\DD}{{\cal D}}
\newcommand{\ZZ}{{\cal Z}}
\newcommand{\Gg}{{\sf G}}

\newcommand{\VV}{{\cal V}}

\newcommand{\tr}{\mbox{tr}}
\newcommand{\Tr}{\mbox{Tr}}
\newcommand{\be}{\begin{equation}}
\newcommand{\ee}{\end{equation}}
\newcommand{\beq}{\begin{eqnarray}}
\newcommand{\eeq}{\end{eqnarray}}
\newcommand{\ba}{\begin{eqnarray}}
\newcommand{\ea}{\end{eqnarray}}
\newcommand{\bse}{\begin{subequations}}
\newcommand{\ese}{\end{subequations}}
\newcommand{\ds}{\displaystyle}
\newcommand{\im}{\mbox{Im}}
\newcommand{\ts}{\textstyle}

\begin{document}
\title{Theory of heterogeneous viscoelasticity}
\author{
Walter Schirmacher$^{1,2,3}$, 
Giancarlo~Ruocco$^1$, and
Valerio Mazzone$^1$,
}
\affiliation{%
$^1$Dipartimento di Fisica, Universit\'a di Roma
``La Sapienza'', P'le Aldo Moro 2, I-00185, Roma, Italy,\\
$^2$Institut f\"ur Physik, Universit\"at Mainz, Staudinger Weg 7,
D-55099 Mainz, Germany,\\
$^3$Institut f\"ur Theoretische Physik,
Leopold-Franzens-Universit\"at Innsbruck, Technikerstra{\ss}e 25/2,
A-6020  Innsbruck, Austria}

\begin{abstract}
{\parindent0em
We review a new theory of viscoelasticity of a
glass-forming viscous liquid near and below the glass transition. 
In our model we assume that each point
in the material has a specific viscosity, which varies randomly
in space according to a fluctuating activation free energy. We include
a Maxwellian elastic term and assume that the corresponding shear
modulus fluctuates as well with the same distribution as that
of the activation barriers.
The model
is solved in coherent-potential
approximation (CPA), for which 
a derivation is given. 
The theory predicts an Arrhenius-type temperature
dependence of the viscosity in the vanishing-frequency limit,
independent of the distribution of the activation barriers.
The theory implies that
this activation energy is generally different from that of a diffusing
particle with the same barrier-height distribution. 
If the distribution of activation barriers  is assumed to have Gaussian
form, the finite-frequency version of the theory
describes well the typical low-temperature
alpha relaxation peak of glasses. Beta relaxation can be included by
adding another Gaussian with center at much lower energies than that
responsible for the alpha relaxation.
At high frequencies our theory
reduces to the description of an elastic medium with spatially fluctuating
elastic moduli
(heterogeneous elasticity theory), which explains the occurrence
of the boson-peak-related vibrational anomalies of glasses.
}
\pacs{65.60.+a}
\end{abstract}
\maketitle
\section{Introduction}

\maketitle
It has been shown by \textcite{maxwell67} that in materials with
high viscosity the elastic response becomes as important as the
viscosity. He suggested that the shear rate of such materials
is given by the sum of a viscous term $\sigma/\eta$ (where $\sigma$
is the applied stress and $\eta$ the viscosity) and an elastic
term $\frac{d}{dt}\sigma/G_\infty$ (where $G_\infty$ is the high-
frequency elastic modulus). This suggestion implies 
- in agreement with experimental findings -
that a high-viscosity
material acts as an elastic material if an alternating stress with high
enough frequency $\nu=\omega/2\pi$ is applied. As shown by
\textcite{maxwell67} such a material tends exponentially
to thermal equilibrium
after an external shear perturbation with a relaxation time
\be\label{maxwell}
\tau=\eta/G_\infty\, .
\ee
If the time scale of external forces is smaller
than $\tau$ the material acts like a solid, if it is larger than
$\tau$ like a liquid. Therefore one defines the glass-transition
temperature $T_g$ to be that temperature at which $\tau$ is larger
than a typical time scale of a glass blower's manipulation, which
corresponds to a viscosity of $\sim 10^{12}$ Pa s.

In glass-forming materials $\eta$ varies exponentially with the inverse
temperature as $\eta(T)\propto \exp\{E_A(T)/k_BT\}$. If the differential
activation energy $E_A(T)$ does not depend on temperature, the material
is called {\it strong}, if it does, {\it fragile}
\cite{angell85,angell91,angell01}. In fragile materials $E_A$ is often parametrized
with the Vogel-Fulcher equation $E_A/k_BT=B/(T-T_0)$, which would lead
to a divergence at $T_0$. Before reaching this value, however,
$E_A(T)$ becomes constant in most substances.

The value of this low-temperature activation energy of the viscosity 
does not agree to that of the diffusivity ({\it Stokes-Einstein
violation}) \cite{fujara92,hodgdon93,chen06}. We shall give an explanation
for this anomaly below.
 
It has also been noticed,
that the activation energy $E_A(T)$ has a similar temperature
dependence as the high-frequency shear modulus
$G_\infty(T)$
\cite{dyre96,dyre98,dyre06,buchenau09,dyre12,hecksher15}.
This has been rationalized by observing that the activation
barrier, which has to be overcome during an activated
relaxation step, comprises essentially elastic energy, which
only involves the shear degrees of freedom \cite{dyre98}, 
and one writes (``shoving model'' \cite{dyre96,dyre98,dyre06,buchenau09,dyre12,hecksher15})
\be
E_A=VG_{\infty}\,.
\ee
The proportionality factor $V$
is the volume of the material region, which participates in the
relaxation step (activation volume). This relation will be of
significance for our model of heterogeneous viscoelasticity
to be described below.

Following \textcite{maxwell67}
one can define a frequency-dependent effective viscosity as
\be\label{maxwell1}
\frac{1}{\eta_{\rm eff}(T,\omega)}=\frac{1}{\eta(T)}+\frac{s}{G_\infty}\, ,
\ee
where
\mbox{$s=-i\omega+\epsilon$} is the Laplace frequency parameter (with
\mbox{$\epsilon\rightarrow$ +0).} This quantity is related to an effective
frequency-dependent shear modulus 
\mbox{$G_{\rm eff}(\omega)=s\eta_{\rm eff}(\omega)=G'(\omega)-iG''(\omega)$}.
Using Eqs. 
(\ref{maxwell}) 
and
(\ref{maxwell1}) 
one obtains
for the mechanical loss modulus the Debye-shaped function
\be
G''(\omega)=G_\infty\frac{\omega\tau}{1+(\omega\tau)^2}
\ee
The maximum 
of the loss function, which is 
due to the structural relaxation - characterized by Maxwell's
relaxation time $\tau$ - has been called $\alpha$ maximum and the process
$\alpha$ (primary) relaxation. The maximum of dielectric 
loss function, which is coupled
to the mechanical relaxation 
\cite{gemant35,dimarzio74,niss05}, 
shows also the temperature variation of $\eta(T)^{-1}$,
but the shape of the dielectric spectrum is different from that
of the mechanical spectrum
\cite{niss05,niss05a}.

However, in all glass-forming materials the $\alpha$
maximum of mechanic and dielectric loss peaks deviates strongly
from the Debye shape, i.e. it is much broader
({\it stretching}). Furthermore,
on the high-frequency (or low-temperature) side of the alpha
peak there is a second feature, which in some materials 
is a peak, in others only a shoulder: the secondary or beta relaxation peak 
\cite{johari70,lunk08,yu13}.
It is commonly believed that, while
the alpha peak describes viscous structural relaxation, the beta peak
is due to residual motions within the already frozen glassy
material \cite{lunk08,yu13}.

Above $T_g$ many features of the glass transition, in particular the
associated non-trivial fractal time dependence of the relaxation functions
- including the stretching of the $\alpha$ relaxation peak -
are captured by the mode-coupling theory (MCT) 
\cite{bengtzelius84,gotze91,gotze09}. This theory 
- in its original version -
describes a sharp transition
towards a non-ergodic state, in which the relaxation function does not fully
decay, but tends to a finite value $f$, the non-ergodicity parameter.
This transition appears at a transition temperature $T_c$, which is
higher than $T_g$. 

It has become clear in the meantime that the critical temperature
of the MCT denotes
not a sharp transition, but rather
a crossover in the liquid dynamics from a fluid
regime to an activated regime \cite{parisi10,berthier11}. 
The activated dynamics is missing in the original MCT.

In
order to take the activated dynamics into account MCT has therefore been
generalized 
\cite{gotzesjogren87,sjogren90,fuchs92,chong08}.
In this version MCT does not only treat momentum relaxation, 
which can be visualized by a succession of scattering events
but also density relaxation, which is a succession of
hopping events (relaxation steps). Such a distinction is also
important in the theory of high-resistivity metals
\cite{belitz83}, where it leads to a cross-over
in the temperature coefficient of the electronic resistivity.

Within such a treatment 
\cite{gotzesjogren87,sjogren90,fuchs92,chong08,belitz83}
the resistive transport coefficient -
in our case the effective viscosity - is given by
\be\label{mct}
\frac{1}{\eta_{\rm eff}}(s)\propto  \delta(T)+\frac{1}{m(s)}\,,
\ee
where $\delta(T)\propto \exp\{-E_A/k_BT\}$ 
describes the activated
viscous motion
and $m(z)$ is the MCT
memory function, which 
generalizes the scattering and contains the mechanism of
structural arrest.
If one replaces the memory function
by the result for the ideal-glass regime of the original MCT, namely
$m(z)=m_f/s$, where $m_f$ is the long-time limit of the memory function,
which can be identified with the shear modulus of the glass
\cite{franosch98,chong06}, one arrives at Maxwell's conjecture
(\ref{maxwell1}).

On the other hand, there is nowadays ample evidence, in particular from
molecular-dynamics simulations, that near $T_g$ the local re-arrangements
exhibits a strong spatial heterogeneity 
\cite{berthier11,berthier11a,berthier11b}. 
({\it dynamical heterogeneity}).

There are presently continuous
efforts to formulate a theory of the glass transition which combines
dynamical heterogeneity
and the mode-coupling scenario
\cite{kirkpatrick87,bouchaud04,parisi10,%
berthier11,szamel10,franz11,franz11a,franz12}.
These theories make contact to former spin-glass theories and
the related replica formalism. In particular
aspects of dynamical heterogeneity have been recently 
attempted to be incorporated
into MCT by treating it as a Landau-type mean-field approximation
within the replica approach
and introducing the Gaussian fluctuations beyond the mean-field
saddle point \cite{parisirizzo13,rizzo14,rizzovoigt14,rizzovoigt15}.

The idea of  heterogeneity of relaxation processes has been also developed
together with the concept of a very ragged free-energy landscape
in configuration space, in which the glassy relaxation is
considered to take place \cite{goldstein69,johari70,stillinger95,stillinger01}.

In many publications
dealing with dielectric and mechanic loss measurements 
\cite{gilroy81,richert93,richert97,buchenau09a}
one quantifies the ideas of dynamic heterogeneity and
ragged free-energy landscape to assume that the Maxwell-type
relaxation processes may take place independently and one would be
allowed to take the average over a distribution of relaxation
barriers $E_i$ or relaxation times $\tau=\tau_0e^{E_i/k_BT}$,
\be\label{gilroy}
G''(\omega)=G_\infty\int d\tau g(\tau)\frac{\omega\tau}{1+(\omega\tau)^2}\,.
\ee
Many authors, e. g.
\textcite{goldstein69} and \textcite{palmer84}, however,
point out that near the glass
transition relaxation processes are highly cooperative and are likely
to involve more events that occur in series than in parallel.
In fact, it is known \cite{kirkpatrick73},
that spatially heterogeneous transport is neither well described by
a parallel nor serial equivalent circuit. This is so, because the
currents seek the path of least resistance, which, in the strong-disorder
limit, amounts to a percolation problem, which is well described by
coherent-potential and effective medium approximations
\cite{bruggeman35,kirkpatrick73,elliott74,kohler13}. 

The phenomenon of dynamic heterogeneity in glass-forming
materials is paralleled by the observation that 
in the high-viscosity/glassy regime the local elastic
moduli also exhibit spatial heterogeneities
\cite{leonforte5a,leonforte6a,wagner11,marruzzo13,mizuno13,mizuno13a,fan14,mizuno14}
({\it elastic heterogeneity}). These, in turn, can be shown to be responsible 
for vibrational anomalies,
which trade under the name ``boson peak''
\cite{schirm98,schirm06,schirm07,zorn11,marruzzo13,kohler13,schirm13,schirm14}.
Quite recently, the present authors have demonstrated \cite{schirm15}, that the two
types of dynamic heterogeneities can be reconciled within the same
theoretical framework, namely the coherent-potential approximation
(CPA)
\cite{kohler13}, applied
to spatially inhomogeneous viscoelasticity. 
In the present contribution
we review this theory and, in particular, show, how it can be generalized
to include also secondary relaxation. 

In the next section a derivation of the CPA for a spatially
fluctuating viscosity is derived. In section III. the heterogeneous
model is introduced and solved. The consequences for the
alpha and beta relaxation and
the $\omega=0$ viscosity
are demonstrated
and discussed. The paper is finished by some conclusions.

\section{Derivation of a CPA
for a spatially fluctuating viscosity}
This derivation follows closely the derivation of the CPA
for heterogeneous-elasticity theory, i.e. elasticity
theory with spatially fluctuating elastic coefficients
\cite{kohler13}.

We consider a heterogeneous viscous liquid in which the
viscosity has different values at different locations in space,
$\eta\rightarrow\eta(\rr)$
which are assumed to fluctuate according to a given distribution
$P[\eta(\rr)]$. Representing the pressure term as usual in terms
of the compressibility $K$, the linearized Navier-Stokes equations
in frequency space take the form \cite{hansen,schirmacher}
\mbox{($s=-i\omega+\epsilon$)}:
\be\label{navierstokes}
{s\rho_m}v_\ell(\rr,s)=
\sum_m\bigg(\frac{K}{s}\partial_\ell\partial_mv_m(\rr,s)
+2\partial_m \eta(\rr)\hat \VV_{\ell m}(\rr,s)\bigg)
\ee
where $\rho_m$ is the mass density and
$\partial_\ell\equiv\partial/\partial x_\ell$.
$v_\ell(\rr,s)$ are
the Cartesian coefficients 
of the Eulerian velocity
field,
and $\widehat\VV$
is the traceless shear strain rate tensor
\mbox{$\widehat \VV_{\ell m}
=\VV_{\ell m}
-\frac{1}{3}\tr\big\{\VV\delta_{\ell m}\big\}$}
with
$\VV_{\ell m}=\frac{1}{2}\big(\partial_\ell v_m+\partial_mv_\ell\big)$.

We can cast this set of equations into the form
\be\label{navier1}
0=\sum_mA_{\ell m}[\eta]
\ee
where the linear operator $A[\eta]$ has the matrix elements
\be
<\rr|A[\eta]|\rr'>_{\ell m}=A_{\ell m}[\eta]\delta(\rr-\rr')
\ee
with
\ba\label{matrix2}
A_{\ell m}&=&s\delta_{\ell m}-\frac{1}{\rho_m}\bigg(\frac{K}{s}\partial_\ell\partial_m
-\frac{2}{3}\partial_\ell \eta(\rr)\partial_m\\
&&+\partial_m \eta(\rr)\partial_\ell+\delta_{\ell m}
\sum_n\partial_n \eta(\rr)\partial_n\bigg)\nonumber
\ea
$A$ is the inverse of the resolvent operator of the linear stochastic
equations (\ref{navierstokes}). Correspondingly, the matrix of
Green's function \footnote{We denote the Green's function
with a {\it sans serif} font $\Gg$ in order to
distinguish it from the shear modulus $G$.} is given by
\be
\Gg(\rr,\rr')_{\ell m}=<\rr|A^{-1}[\eta]|\rr'>_{\ell m}
\ee
This matrix can be represented as a functional integral over
mutually complex-conjugate vector fields 
\cite{john83,kohler13}
$v_\ell^\alpha(\rr)$, $\bar v_m^\alpha(\rr)$
in $n$ replicas of the system\footnote{At the end of the
calculation one has to take $n\rightarrow 0$.}
($\alpha=1,\dots,n$)
as
\ba
\Gg(\rr,\rr')_{\ell m}
&=&\prod\limits_{\alpha=1}^n\prod\limits_{\mu \nu}\int
\DD[\bar v^\alpha_\mu(\rr),
v^\alpha_\nu(\rr)]
\bar v^1_\ell(\rr)
v^1_m(\rr')\nonumber\\
&&\times\,\eee^{\ts -\sum\limits_{\alpha}<\vv^\alpha|A|\vv^\alpha>}\\
&=&\frac{\delta}{
\delta J^{(1)}_{\ell m}(\rr,\rr')
}\ZZ[J(\rr,\rr')]\bigg|_{J=0}
\ea
Here we have defined a generating functional
\ba
&&\ZZ[J(\rr,\rr')]
=\prod\limits_{\alpha=1}^n\prod\limits_{\ell m}\int
\DD[\bar v^\alpha_\ell(\rr),
v^\alpha_m(\rr)]\nonumber\\
&&\times\,\eee^{\ts -\sum\limits_{\alpha}<\vv^\alpha|A|\vv^\alpha>}
\eee^{\ts -\sum\limits_{\alpha}<\vv^\alpha|J^\alpha|\vv^\alpha>}
\ea
\mbox{with source-field matrices $J^\alpha_{\ell m}(\rr,\rr')$.}
The operator $A$ can now be identified as the action of a Gaussian
field theory. Its matrix element can be cast into the form
\ba
&&<\vv^\alpha|A|\vv^\alpha>=
\int d^3\rr \frac{1}{\rho_m}\bigg(
s\sum_\ell
|v_\ell^\alpha(\rr)|^2\nonumber\\
&&+\frac{1}{2s} K\,\tr\{\VV^\alpha(\rr)\}^2
+
\eta(\rr)\sum_{\ell m}|\widehat\VV^\alpha_{\ell m}(\rr)|^2\bigg)
\ea
where the trace $\tr$ runs over the Cartesian indices.
We now apply a procedure common in deriving effective
field theories (Fadeev-Popov method \cite{belitz97}).
First we replace the fluctuating viscosity $\eta(\rr)$ by
a ``place-holder'' field $Q^{(\alpha)}(\rr,s)$ with the help
of a functional delta function. The latter is then, in turn,
expressed as a functional integral over a second auxiliar
field $\Lambda^{(\alpha)}(\rr,s)$:

\ba\label{}
\ZZ[J]&=&
\int\mathcal{D}[\vv,\bar{\vv}]\,\int\mathcal{D}[Q]\,\eee^{-<\vv|A[Q]-J|\vv>}\delta[\eta-Q]\nonumber\\
&=&\int\mathcal{D}[\vv,\bar \vv]\,\mathcal{D}[Q,\Lambda]\,\eee^{-<\vv|A[Q]-J|\vv>}\eee^{<\Lambda|\eta-Q>}\nonumber\\
&=&\int\mathcal{D}[Q,\Lambda]\,\eee^{-\Tr\{\ts\,\ln[\,\Aa[Q]-J\,]\,\}}\eee^{<\Lambda|\eta-Q>}\label{fadeev}
\ea

In this expression we have
suppressed the replica indices for
brevity. 
In the third line of Eq. (\ref{fadeev}) we have integrated out
the original velocity fields
$\bar \vv^\alpha$ and $\vv^\alpha$. This can be done, because the
functional integral in the second line is just a Gaussian.
The trace $\Tr$ has to be taken over the
continuous spatial variables, the Cartesian indices and
the replica indices.

We now follow \textcite{kohler13} in coarse-graining our system
of volume $V$
into $N_c=V/V_c$ cells of diameter $\xi$ and volume
$V_c=\xi^3$, which is the correlation length of
the fluctuations $\Delta\eta(\rr)=\eta(r)-\langle\eta\rangle$, defined
by
\be
\xi^3=\frac{1}{\langle\eta^2\rangle}\int d^3\rr
\langle\Delta\eta(\rr+\rr_0)\Delta\eta(\rr_0)\rangle
\ee
The fluctuating field $\eta(\rr)$ is then averaged over  a given
cell with label $i$, which gives a value $\eta_i$, which still
fluctuates from cell to cell. 
The statistical fluctuations of 
these values can now be assumed to
be uncorrelated, i.e.
\be
P(\eta_1\dots \eta_i\dots \eta_{N_c})
=\prod\limits_{i=1}^{N_c}p(\eta_i)
\ee
We now associate with all fields discretized numbers corresponding
to the center vectors $\rr_i$ of the cells:
$\Lambda^{(\alpha)}(\rr)\rightarrow\Lambda^{(\alpha)}_i$ and
$Q^{(\alpha)}(\rr)\rightarrow Q^{(\alpha)}_i$.
Using this the scalar product, which appears in the
exponential in Eq.  
(\ref{fadeev}), can be written as:

\be
<\Lambda|\eta-Q>
=\frac{V_c}{V}\sum_{\alpha}\sum_i\Lambda_i^{(\alpha)}(\bm{r})\left(\eta_i^{(\alpha)}-Q_i^{(\alpha)}\right)
\ee
We now start to evaluate the configurational average.
Due to the Fadeev-Popov transformation the only
term to be averaged over is the term 
$e^{<\Lambda|\eta-Q>}$.

Assuming that all the $N_c$ coarse-graining cubes behave the 
same on average and using that the 
individual cubes are not correlated, we can write

\begin{eqnarray}
&\left\langle e^{<\Lambda|\eta-Q>}\right\rangle
=\prod_\alpha \prod_i
\left\langle \eee^{\ts\frac{V_c}{V}\Lambda_i^{(\alpha)}(\eta_i^{(\alpha)}-Q_i^{(\alpha)})}\right\rangle_i \nonumber&\\
&=\eee^{\sum_{\alpha}\frac{V}{V_c}\ln\left(\left\langle \exp\left[\,-\frac{V_c}{V}\Lambda_i^{(\alpha)}(\eta_i^{(\alpha)}-Q_i^{(\alpha)})\,\right] \,\right\rangle_i\,\right)} \label{avCGVolume}&
\end{eqnarray}

Note that the two occurring volume ratios do not cancel each other due to the average inside the logarithm.
Using (\ref{avCGVolume}) the generating
functional (\ref{fadeev}) can be written as

\be\label{final}
\ZZ[\tilde J]=\int\mathcal{D}[Q,\Lambda]\, \eee^{\ts-S_{\text{eff}}[Q,\Lambda,\tilde J]}
\ee
where we have now replaced the source field matrix
$J^\alpha(\rr,\rr')$ by translational-invariant one
$\tilde J(\rr-\rr')$, which is  not supposed
to depend on the replica index $\alpha$.
The effective action takes the form
\ba\label{skalphonseff}
S_{\text{eff}}[Q,\Lambda,\tilde J]&=&\Tr\{\,\ln\big(\Aa[Q]-\tilde J\big)\}\\
&&-\displaystyle\sum_{\alpha=1}^n\frac{V}{V_c}\ln\left(\left\langle \eee^{-\frac{V_c}{V}\Lambda_i^{(\alpha)}(\eta_i^{(\alpha)}-Q_i^{(\alpha)})} \,\right\rangle_i\,\right)\nonumber
\ea
Since the factor $\frac{V}{V_c}$ in the effective action (\ref{skalphonseff}) is much larger than unity a saddle point approximation can be employed to evaluate the integral in (\ref{final}). 

We now assume the ``fields
$Q_i^{(\alpha)}$ and
$\Lambda_i^{(\alpha)}$
to be the same in all replicas.
Varying the effective action with respect to the fields
$Q_{i,s}$ and $\Lambda_{i,s}$
yields the following equation for the saddle-point quantities
$Q_{i,s}^{(\alpha)}$ and
$\Lambda_{i,s}^{(\alpha)}$:

\ba
 0&=&\frac{\left\langle-\frac{V_c}{V}\Lambda_{i,s}^{(\alpha)}(\eta_i-Q_{i,s})\eee^{-\frac{V_c}{V}\Lambda_i(\eta_i-Q_{i,s})}\right\rangle_i}{\left\langle \eee^{-\frac{V_c}{V}\Lambda_{i,s}(\eta_i-Q_{i,s})}\right\rangle_i}\, , \nonumber
\ea
from which follows
\ba
0&=&\left\langle\frac{\eta_i-Q_{i,s}}{\exp[\frac{V_c}{V}\Lambda_{i,s}(\eta_i-Q_{i,s})]}\right\rangle_i\label{saddleb}
\ea

Since $\frac{V_c}{V}\ll1$ the exponential in the denominator can be expanded to first order:

\begin{subequations}
\be\label{cpa1a}
0=\left\langle\frac{\eta_i-Q_{i,s}}{1+\frac{V_c}{V}(\eta_i-Q_{i,s})\Lambda_{i,s}}\right\rangle_i
\ee

The second saddle point equation gives
\ba\label{cpa1b}
\left.\frac{\partial\,\tr\{\,\ln[{A}(Q)\,]\,\}}{\partial Q_i}\right|_{Q_i=Q_{i,s}}&=&\nonumber
\frac{\frac{V_c}{V}\Lambda_{i,s}\left\langle\eee^{-\frac{V_c}{V}\Lambda_i(\eta_i-Q_{i,s}} \right\rangle_i}{\left\langle\eee^{-\frac{V_c}{V}\Lambda_{i,s}(\eta_i-Q_{i,s}} \right\rangle_i}\\&=&\frac{V_c}{V}\Lambda_i
\ea
\end{subequations}
The left-hand side can be evaluated under the assumption that the saddle point field $Q_s$ is constant in space, i.e.
$Q_{i,s}\equiv Q$ for all $i$.
This corresponds to the introduction of an effective homogeneous medium
in which 
\be
Q(s)=\eta(s)=G(s)/s
\ee
As in all effective-medium theories the effective medium is identified
with the real medium, in which $\eta(s)$ is the macroscopic
frequency-dependent
viscosity and $G(s)$ the corresponding macroscipic shear modulus.

The homogeneus Matrix $A[Q]$ is both diagonal in the Cartesian 
indices and with respect to the $\kk$ vectors in $\kk$ space. 
The three diagonal elements (in Cartesian space)
are the inverse longitudinal and transverse Green's functions
of the effective medium
$1/\Gg_L(\kk, s)$ and (two entries) $1/\Gg_T(\kk, s)$, which are given by
\begin{subequations}
\be
1/\Gg_L(\kk, s)=s+\frac{1}{\rho_m}k^2\bigg(\frac{K}{s}+\frac{4}{3}Q(s)\bigg)
\ee
\be
1/\Gg_T(\kk,s)=s+\frac{1}{\rho_m}k^2Q(s)
\ee
\end{subequations}
Defining a new field (``susceptibility function'')
\mbox{$\widetilde\Lambda(s)=3V_c/\widetilde{\nu} V\Lambda^\alpha(s)$}
with $\widetilde{\nu}=\nu^3/2\pi^2$
the CPA equations (\ref{cpa1a}) and (\ref{cpa1b}) become:
\begin{subequations}
\be\label{cpa2a}
0=\left\langle\frac{\eta_i-Q(s)}{1+\frac{\widetilde \nu}{3}(\eta_i-Q(s))
\widetilde\Lambda(s)}\right\rangle_i
\ee
\be\label{cpa2b}
\widetilde\Lambda(s)=
\frac{3}{k_\xi^3}\int_0^{k_\xi}dk k^4\frac{1}{\rho_m}\bigg(\frac{4}{3} \Gg_L(k,s)+2\Gg_T(k,s)\bigg)
\ee
\end{subequations}
It is easily shown that the CPA equation (\ref{cpa2a}) is equivalent to
the following equations
\begin{subequations}
\be\label{cpa2c}
Q(s)=\left\langle\frac{\eta_i}{1+\frac{\widetilde \nu}{3}(\eta_i-Q(s))
\widetilde\Lambda(s)}\right\rangle_i
\ee
\be\label{cpa2d}
1=\left\langle\frac{1}{1+\frac{\widetilde \nu}{3}(\eta_i-Q(s))
\widetilde\Lambda(s)}\right\rangle_i
\ee
\end{subequations}
For frequencies, which are much smaller than those, in which
the inertial term in the Navier-Stokes equation (\ref{navierstokes})
is important (at GHz frequencies and above) the susceptibility function
$\widetilde\Lambda(s)$ can be replaced by its low-frequency limit
\be\label{lowfrequency}
\widetilde\Lambda(s)
\stackrel{s\rightarrow 0}{\longrightarrow}\,\,
\frac{2}{Q(s)}
\ee
\section{Heterogeneous-viscoelasticity theory}
\subsection{Model}
We now consider a heterogeneous viscoelastic model, in which
equation (\ref{maxwell1}) holds locally with both spatially
fluctuating viscosity and shear modulus
\be\label{heteromaxwell}
\frac{\ds 1}{\ds \eta_{\rm eff}(\rr,s)}
=
\frac{\ds 1}{\ds \eta(\rr)}+\frac{\ds s}{\ds G(\rr)}
\ee
The local viscosity is assumed to be governed by
a local free energy
$\ln[\eta(\rr)/\eta_0]=F(\rr)/k_BT$ with
$F(\rr)=E(\rr)-TS(\rr)$.
$E$ is the local energy barrier and $S$ is a multi-excitation
entropy \cite{yelon90,yelon06}, which is related to $E$ by
a compensation
(Meyer-Neldel) rule \cite{keyes58,lawson60,yelon90,yelon06}
$S(\rr)/k_B=\alpha E(\rr)$, so that we have
$\eta(\rr)=\eta_0e^{\beta_{\rm eff}E(\rr)}$
with $\beta_{\rm eff}=[k_BT]^{-1}-\alpha$. The activation barrier, in turn,
is assumed \cite{dyre06,hecksher15} to be related by $E(\rr)=VG(\rr)$ to the local
high-frequency shear modulus, where $V$ is an activation volume.

Taking (\ref{lowfrequency}) into account the CPA equation (\ref{cpa2c})
becomes
\be\label{cpa3}
Q(s)\equiv\eta(s)=\left\langle\frac{
\eta_{\rm eff}^{(i)}
}{1-\frac{2\widetilde\nu}{3}+\frac{2\widetilde \nu}{3}
\eta_{\rm eff}^{(i)}
/\eta(s)
}\right\rangle_{\!\!\!P(E)}
\ee
where $\langle\dots\rangle_{P(E)}$ denotes an average over the
distribution $P(E)$ of activation energies $E$.
The explicit form of the local Maxwellian
viscosity is
\be
\frac{1}{\eta_{\rm eff}^{(i)}(s)}=
\frac{1}{\eta_{\rm eff}(E,s)}=\frac{1}{\eta_0}e^{-\beta_{\rm eff}E}
+\frac{sV}{E}
\ee
In contrast to the local viscosity, which is, together with 
the distribution $P(E)$ the {\it input} to the CPA calculation, $Q(s)\equiv\eta(s)$
is the macroscopic frequency-dependent viscosity, which is the
{\it output} of the calculation. The macroscopic viscosity is related
to the macroscopic frequency-dependent shear modulus by
\be
s\eta(s)=G(s)=G'(\omega)-iG''(\omega)
\ee
where $G''(\omega)$ is the mechanical loss function.
 
In the present treatment we use for $P(E)$
a single Gaussian ($\alpha$ relaxation)
and a superposition of two Gaussians ($\beta$ relaxation). We cut
off $P(E)$ at $E=0$, i.e. $P(E)=0$ for $E<0$.

We emphasize that in the very high frequency regime, where 
the full susceptibility function $\widetilde \Lambda(s)$,
Eq. (\ref{cpa2b})
has to be taken and where
viscous effects become irrelevant, our theory reduces to
heterogeneous elasticity theory, which describes the high-frequency
anomalies associated with the boson peak
\cite{schirm98,schirm06,schirm07,zorn11,marruzzo13,kohler13,schirm13,schirm14}.
This means that the
present theory describes both dynamical and vibrational heterogeneities.

\subsection{The case $\omega=0$ and the Stokes-Einstein violation}

As we want to compare in the following the behavior of the 
heterogeneous viscosity with 
diffusive single-particle motion in the same energy landscape,
(heterogeneous diffusivity) in the $\omega=0$ limit,
we quote the CPA equations for this problem from \textcite{kohler13}:

\be\label{diff}
D(s)=
\left\langle
\frac{D^{(i)}}
{1+\frac{\widetilde \nu}{3}
\big(D^{(i)}(s)-D(s)\big)
\Lambda_D(s)}
\right\rangle_i
\ee
with
\mbox{$\Lambda_D(s)=\frac{3}{k_\xi^3}\int_0^{k_\xi}dkk^4
[s+D(s)k^2]^{-1}$}
Here $D(s)$ is the dynamic diffusivity and
$D^{(i)}=D_0e^{-\beta_{{\rm eff},D}E^{(i)}}$
are the local diffusivities with $\beta_{{\rm eff},D}
=[k_BT]^{-1}-\alpha_D$ \footnote{The Meyer-Neldel
parameter $\alpha_D$ need not be the same as thas
for the viscosity.}

In the $s=0$ limit
$\Lambda_D\rightarrow 1/D(s=0)\equiv 1/D$, and we obtain for the $\omega=0$
diffusivity the CPA equation
\bse
\be\label{5a}
\frac{\widetilde \nu}{3}=\int_0^\infty dEP(E)
\frac{1}{
\big(\frac{3}{\widetilde\nu}-1\big)
\frac{D}{D^{(i)}}+1
}
\ee
For the viscosity $\eta=Q(0)$, on the other hand, we obtain
from (\ref{cpa3})
\be\label{5b}
\frac{2\widetilde \nu}{3}=\int_0^\infty dEP(E)
\frac{1}{
\big(\frac{3}{2\widetilde\nu}-1\big)\frac{\eta}{\eta^{(i)}}+1
}
\ee
\ese

If the macroscopic viscosity and diffusivity are
parametrized as 
$\eta\propto e^{\beta_{\rm eff}E_A}$,
$D\propto e^{-\beta_{{\rm eff},D}E_{A,D}}$, the integrands
in Eqs. (\ref{5a}) and (\ref{5b}) become step functions 
$\theta(E-E_A)$ and $\theta(E_{A,D}-E)$, resp. 
in the low-temperature limit,
and we arrive at
\be\label{lowtemp}
1-\frac{2\widetilde\nu}{3}=\int_0^{E_A} dE P(E)\qquad
\frac{\widetilde\nu}{3}=\int_0^{E_{A,D}} dE P(E)
\ee

This means that (within CPA) both the diffusivity and viscosity
with spatially fluctuating activation energies acquire an Arrhenius
behavior, independently of the details of $P(E)$.
This result is well known for the diffusivity and
(for charged carries) conductivity in disordered materials
\cite{efros84,kohler13}. It reflects the fact that the carrier look
for a path of minimum resistance through the material, which is
a percolation path. In the percolation theory of hopping conduction
\cite{efros84} the number $\widetilde \nu/3$ is the continuum
percolation threshold, which we now call $p_D$. The analogous
quantity for the viscosity is $p_\eta=1-2p_D$, where the factor 2 can
be traced back to the two transverse cartesian degrees of freedom of
the shear motion \cite{kohler13}. 
So we note the result that except
for the special case $p_D=1/3$ the activation energy for diffusion
and viscosity should be different. The explanation is that the percolation
process for a single-particle and cooperative motion in three dimension
is different. 
If we take for $p_D=\widetilde\nu/3$ the three-dimensional continuum
percolation threshold $\approx 0.3$ we arrive at
$p_\eta=1-2p_D=0.4$. Using Eqs. (\ref{lowtemp}) we arrive for
a Gaussian distribution centered
at $E_1$ with width parameter $\sigma/E_1$ = 0.3 at
$E_{A,D}/E_1$ = 0.843 and $E_A/E_1$ = 0.925, i.e. the ratio
is $E_{A,D}/E_A$ = 0.91. This ratio (Einstein-violation parameter)
depends on dimensionality through $p_D$, but it is non-universal, as
it depends on (and becomes smaller with) the shape of the distribution.

All the above considerations assume that the energy distribution for the
local diffusion coefficient is the same as that for the local viscosity, 
i.e. that locally the Einstein relation holds. Especially in materials
consisting of several diffusing species - like metallic and ionic glasses -
this assumption is certainly unjustified. It would be much more plausible
to take the small-energy Gaussian, responsible for the beta relaxation,
as that distribution for the diffusion. This would be in agreement
with experimental findings relating the beta energy scale with
that of the diffusion in metallic glasses \cite{yu13}.
\subsection{Alpha relaxation}

\begin{figure}
\includegraphics[width=8cm]{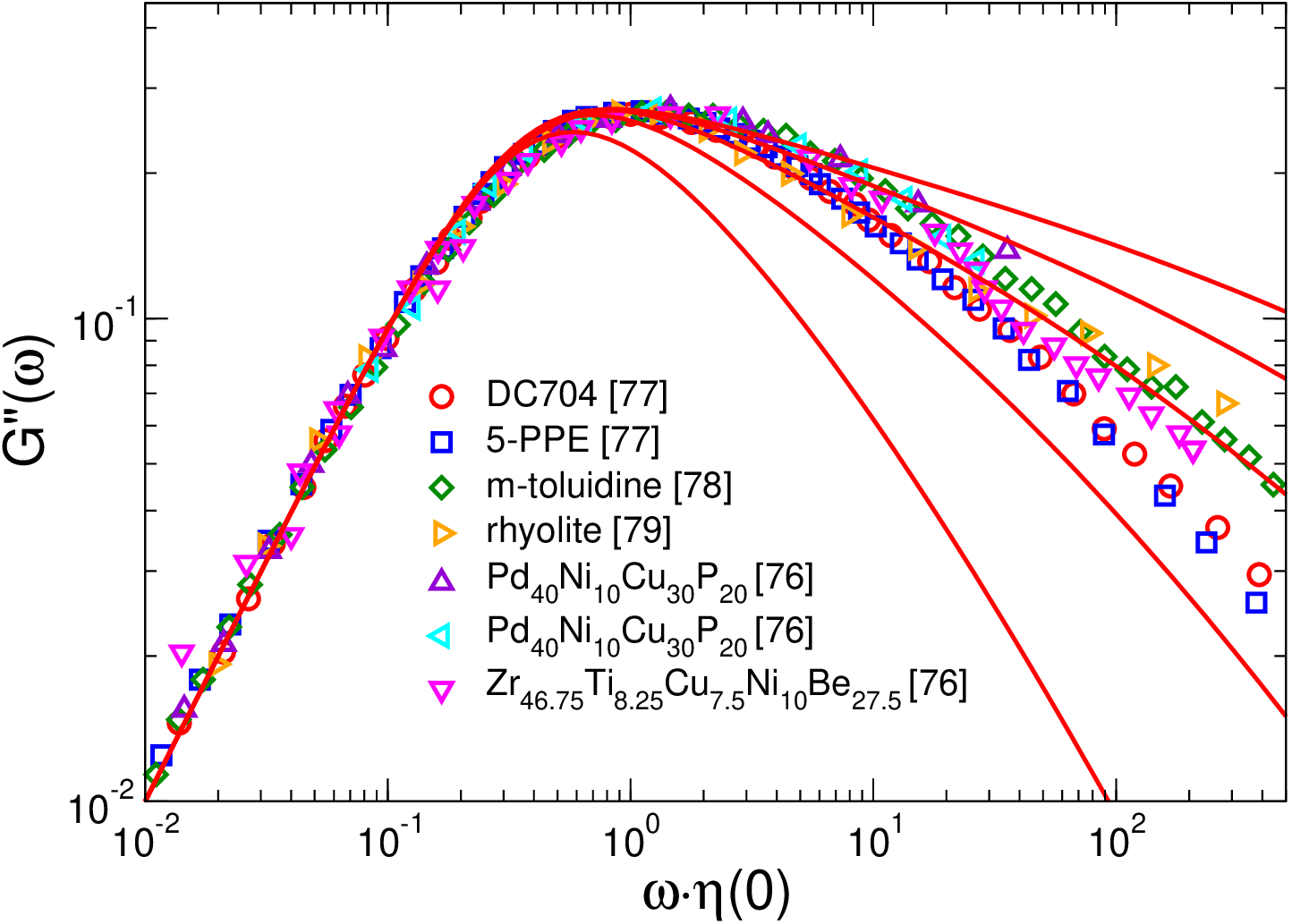}
\caption{
Mechanical loss curves $G''(\omega)=\omega\eta'(\omega)$
vs. $\omega\eta(0)$
for fixed inverse temperature
$\beta_{\rm eff}E_1$ = 26, calculated in CPA according to 
Eq. (\ref{cpa4}) with a single Gaussian distribution
with widths $\sigma$
varied from 0.1 to 0.3 in steps of 0.05.
The experimental data (symbols) are from metallic,
organic and
inorganic glasses 
\cite{wang08,hecksher13,hutcheson08,webb97}. 
}\label{fig1}
\end{figure}

We turn now to the frequency dependence of the viscosity, as
predicted by our theory of heterogeneous viscoelasticity.
We can reformulate the CPA equation (\ref{cpa3}) in terms of
the dimensionless viscosity $\eta(s)/\eta(0)$ \mbox{as follows:}
\be\label{cpa4}
\frac{\eta(s)}{\eta(0)}
=\left\langle
\frac{1}{q_\eta
\left(
e^{\beta_{\rm eff}(E_A-E)}+\frac{sV\eta(0)}{E}
\right)+p_\eta
\frac{\eta(0)}{\eta(s)}
}
\right\rangle_{\!\!\!\!P(E)}
\ee
with $p_\eta=\frac{2\widetilde\nu}{3}$
and $q_\eta=1-p_\eta$.
It can be seen from this representation that the strongest contribution
to the integral
comes from the energy \mbox{$E\approx E_A$,} so that effectively the function
$\eta(s)/\eta(0)$ is approximately a universal function of
the scaled frequency parameter $sV\eta(0)/E_A\approx sV\eta(0)/E_1$,
where $E_1$ is the peak energy of the principal Gaussian
(for $E_A\approx E_1$ see section E). This quasi-universality
is inherited by the loss function 
\mbox{$G''(\omega)\propto\omega\eta(0)\cdot\eta'(\omega)/\eta(0)$}
= \mbox{$\omega \eta'(\omega)$}.
In all our calculations we measure viscosities in units of $\eta_0$,
mechanical loss functions in units of $E_1/\eta_0$ and frequencies
in units of $E_1/V\eta_0$. The activation volume obviously
enters only into
the frequency scale.

\begin{figure}
\includegraphics[width=8cm]{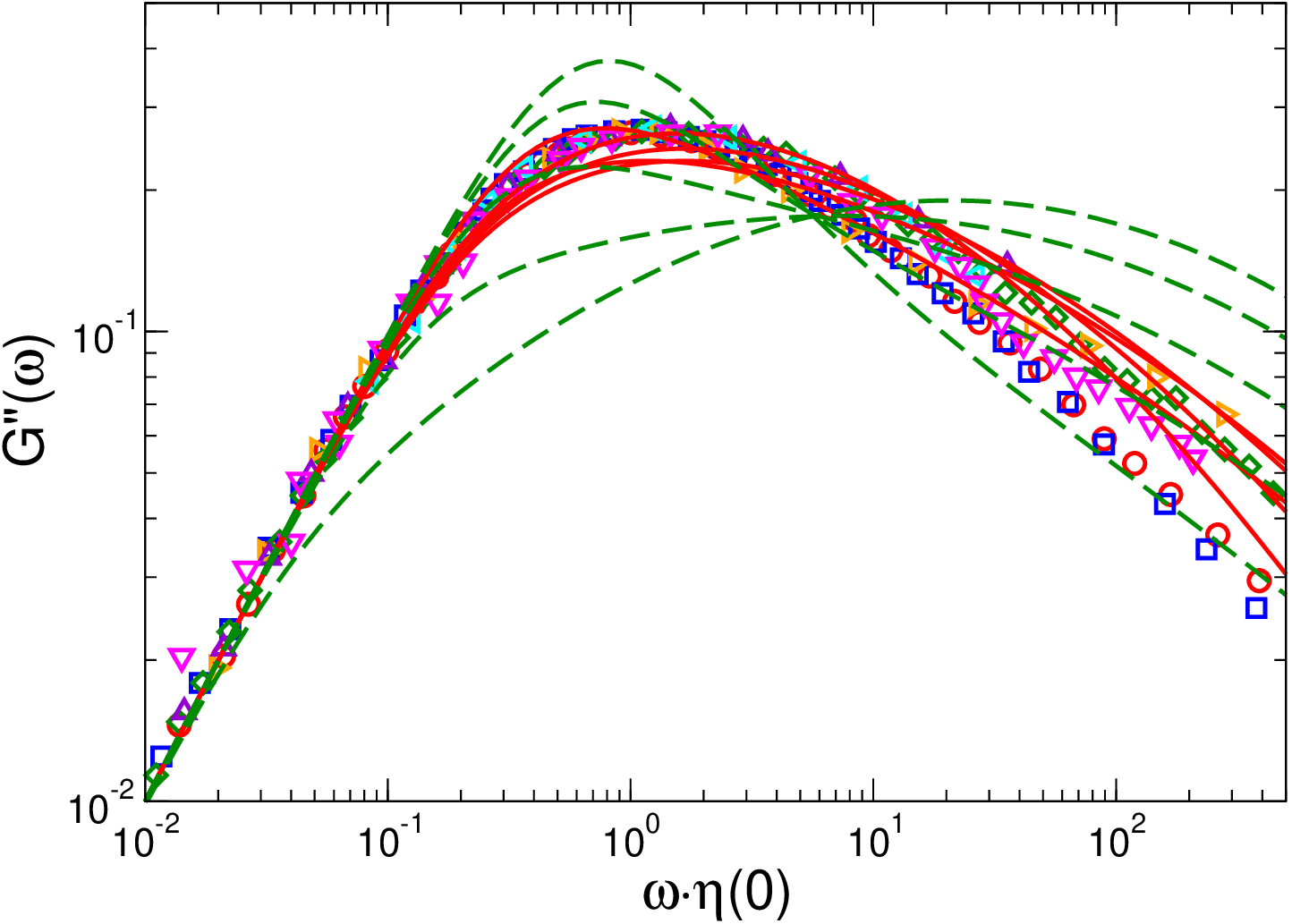}
\caption{
Mechanical loss curves $G''(\omega)=\omega\eta'(\omega)$
vs. $\omega\eta(0)$
for inverse effective temperatures
$\beta_{\rm eff}E_1$ = 16, 18, 20, 22, 24,
calculated in CPA, Eq. (\ref{cpa4})
for a Gaussian distribution
of width $\sigma=0.25$, (straight lines) and in
VCA, Eq. (\ref{vca}), (dashed lines). The symbols denote the same
experimental data as in Fig. \ref{fig1}.}
\label{fig2}
\end{figure}
  
In Fig. \ref{fig1} we show the result
of CPA calculations for a Gaussian distribution of
activation energies with varying width $\sigma$ together with a number
of measured mechanical-loss spectra. 
\cite{wang08,hecksher13,hutcheson08,webb97}%
\footnote{In our calculations we 
measure all energies in units of the center of the
Gaussian distribution
$E_1$}. A very important point is that the left wing of the 
alpha relaxation spectra
is given by $G''(\omega)=\omega\eta(0)$. 
Because previously there was
no proper theory for calculating the $\omega=0$ value of the
viscosity,
efforts to describe the alpha maximum in terms of a distribution
of activation energies have rested on shaky grounds.

Near the alpha peak the frequency dependence of
$\eta(s)$ starts to be effective, which causes the strong asymmetry
of 
the alpha relaxation spectrum.
We see from Fig. \ref{fig1} that for the chosen value
of the inverse effective temperature $\beta_{\rm eff} E_1=24$ the CPA curves for
a width parameter $\sigma/E_1$ = 0.2 fits the data best.

It is of interest to compare our CPA calculation with the previously
used
procedure (Eq. (\ref{gilroy}) ) of averaging the local viscosities.
In the theory of disordered systems this amounts to averaging
over the Hamiltonian and has been called virtual-crystal
approximation (VCA). If the self-consistent term in the
denominator of
(\ref{cpa3}) is omitted,
one obtains
\be\label{vca}
Q(s)=\bigg\langle\eta_{\rm eff}^{(i)}(s)\bigg\rangle\, ,
\ee
which is just the same as (\ref{gilroy}),
except for the fact that in our model $G_\infty$
is also assumed to fluctuate.
 
In Fig. \ref{fig2} we
compare the CPA calculations for different inverse effective
temperatures with VCA calculations using (\ref{vca}). It is clearly
seen that the approximate time-temperature scaling of the alpha
peak, which is obeyed in CPA, is not reproduced in VCA. This can
be traced again to the fact that the $\omega=0$ value of the
viscosity in VCA, which corresponds to a parallel-circuit formula 
is not correct, because it does not recognize the percolation
aspects of the viscous currents. On the other hand, at frequencies
much higher than that of the alpha relaxation peak, the percolation
aspects lose their importance, because at these frequencies the
viscous flow is an alternating one, probing only relaxational
transitions to adjacent free-energy minima. In this frequency regime
- as we will see in the paragraph on beta relaxation - 
the CPA approximately becomes equalent to the VCA,
which means thaat the VCA and the corresponding expressions
(\ref{gilroy}), (\ref{vca}) are justified in the high-frequency/low-temperature
regime.

\subsection{Dielectric relaxation}
\begin{figure}
\includegraphics[width=8cm]{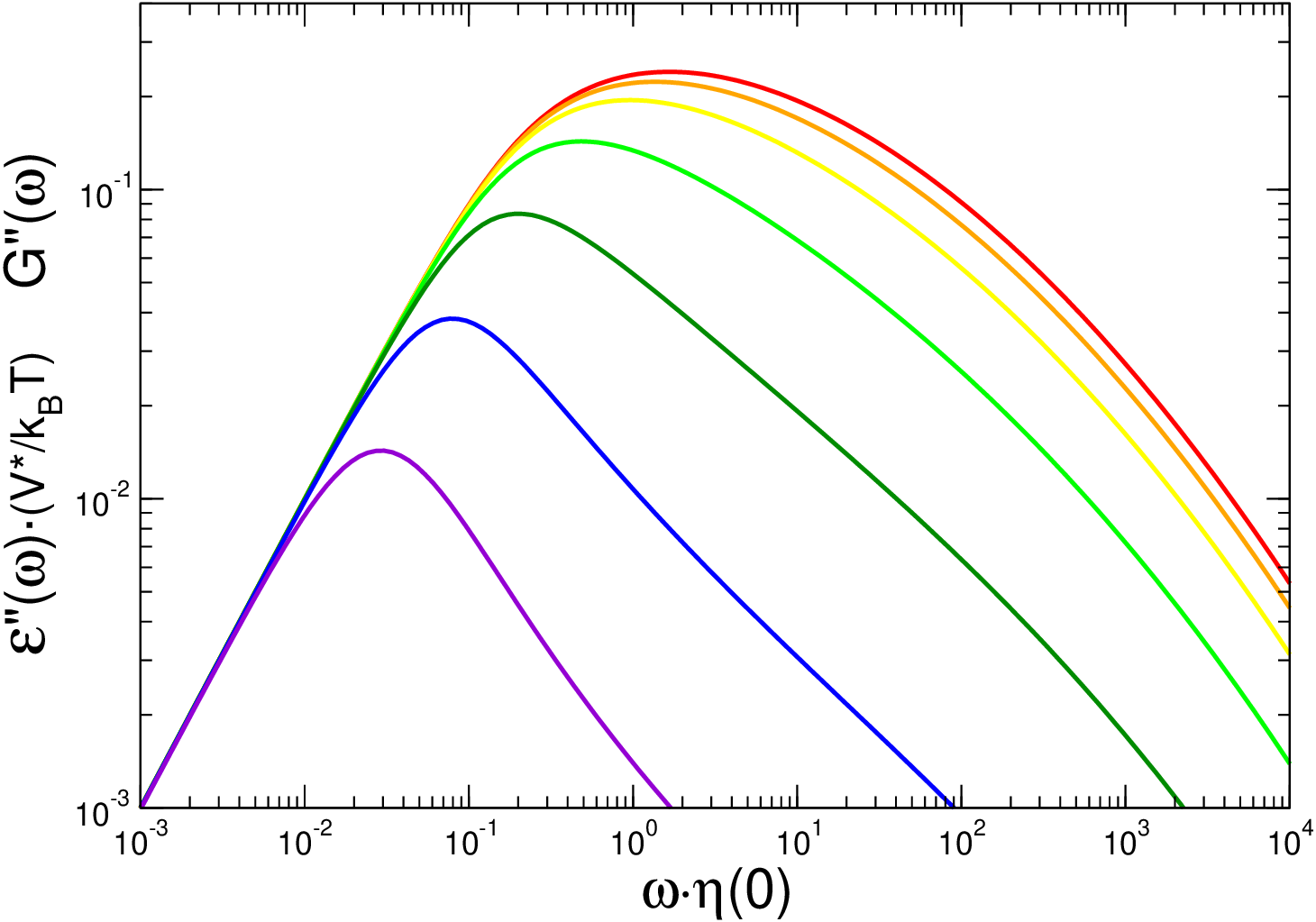}
\caption{Dielectric loss spectra calculated by means of the Gemant-DiMarzio
formula (\ref{genant}) from $\eta(s)$ data calculated in CPA
with $\beta_{\rm eff}E_1$ = 15 and $\sigma/E_1$ = 0.25
and the Gemant-DiMarzio parameter $V^*/k_BT$ varied by decades
(top to bottom)
from $10^{-1}$ to $10^5$. The very top (red) line is the mechanical
loss function $G''(\omega)$.
}
\label{fig3}
\end{figure}
As noted in the introduction the peaks of dielectric loss
data in glass-forming materials follow precisely the
inverse viscosity, and hence the peaks of the mechanical
loss data. It has been noted, however that the shape of
the loss curves are not the same \cite{niss05,niss05a}.
However, according to \textcite{gemant35}, \textcite{dimarzio74}
and \textcite{niss05} they can be related to each other by
\be\label{genant}
\epsilon''(\omega)
=\im\left\{
\frac{1}{1+\frac{V^*}{k_BT}G(s)}
\right\}\,,
\ee
where $V^*$ is a microscopic volume.
A very similar formula can also be obtained within
the mode-coupling formalism, applied to the
coupling of special degrees of freedom
(in this case the local dipoles) to the bulk relaxing
dynamic variables
\cite{sjogren90,franosch97,schilling97},
assuming that the bulk density fluctuations are essentially frozen.

In Fig. \ref{fig3}
we show the result for the dielectric loss for different
values of the coefficient $V^*/k_BT$. It is seen that
with increasing $V^*/k_BT$ the peak is shifted to the left and
the stretching of the alpha peak becomes less pronounced.
A detailed discussion of (\ref{genant}) can be found in
the paper by  \textcite{niss05}.

\subsection{Beta relaxation}
\begin{figure}
\includegraphics[width=8cm]{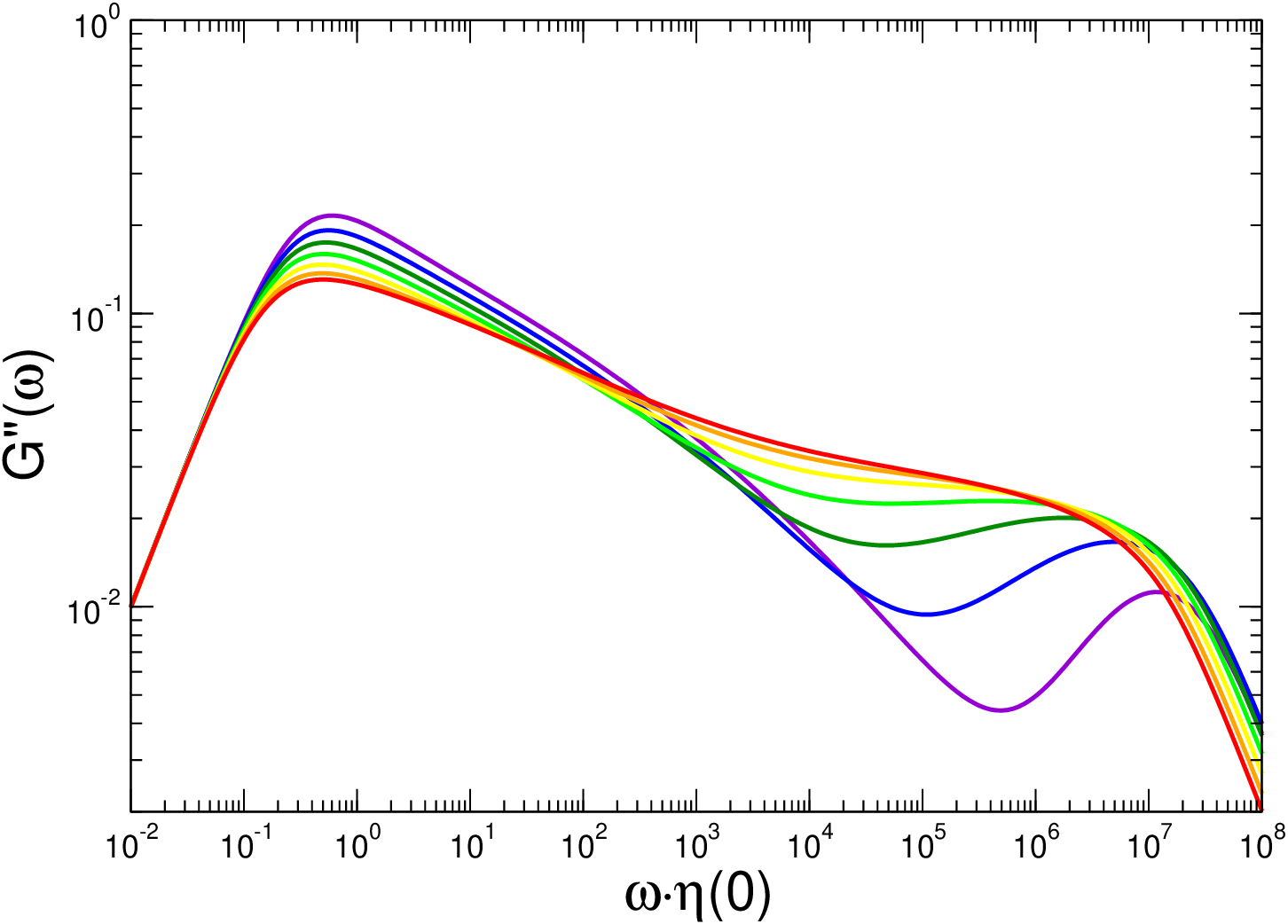}
\caption{%
Mechanical loss spectra for a distribution P(E) with
two Gaussians of weight $x_1$ = 0.8 and $x_2$ = 0.2 centered
at $E_1\equiv 1$ and $E_2/E_1=0.1$. The width of
the high-energy peak is $\sigma_1/E_1=0.25$. The widths 
$\sigma_2/E_1$ of the second,
low-energy peak is varied (from bottom to top) from 0.05 to 0.35
in equal steps of 0.05. The effective inverse temperature is
set $\beta_{\rm eff}E_1$ = 25.}
\label{fig4}
\end{figure}

As noted by \textcite{johari70} and many authors later
\cite{lunk08,yu13} there exists a relaxation regime at the 
high-frequency and low-temperature side of the alpha-relaxation
peak, which is sometime a peak, sometime a shoulder
and sometime just a wing. This part of the energy spectrum has been
reported \cite{yu13} to be related to the glassy, i.e. solid-like
yield dynamics.

We attempted to include this by adding a second Gaussian to the
primary Gaussian, which describes the alpha relaxation, 
but with much lower center.
The distribution is truncated at $E=0$. Therefore with
increasing width of the second Gaussian the latter becomes an almost
constant wing on the low-energy side of the primary Gaussian.
From Fig. \ref{fig4}, where we have varied the width of the secondary
Gaussian, 
we see that by doing this we can describe the transition from a beta
wing to a beta maximum.

\begin{figure}
\includegraphics[width=8cm]{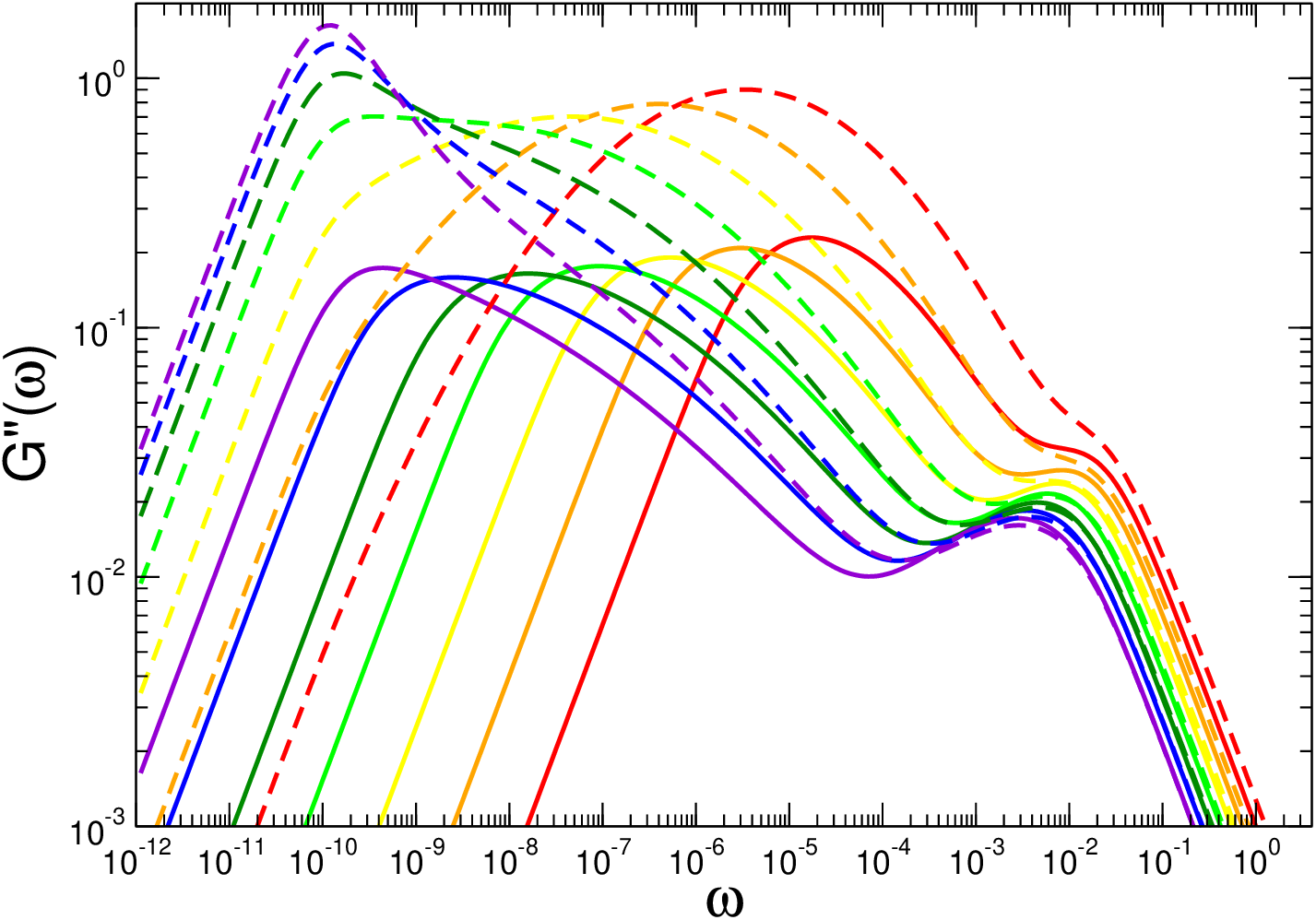}
\caption{
Mechanical loss spectra for a two-Gaussian distribution P(E)
as in Fig. \ref{fig4} calculated in CPA (\ref{cpa3}) and
VCA (\ref{vca}). The parameters are the same as in Fig. \ref{fig4}
except that the 
inverse effective temperature is varied
between $\beta_{\rm eff}E_1$ = 12 and 24, and the width
of the low-energy Gaussian is fixed to be
$\sigma_2/E_1$ = 0.1. Note that the frequency axis is
{\it not} scaled with the $\omega=0$ viscosity.
}
\label{fig5}
\end{figure}

As the beta relaxation is probed at frequencies much higher than the
principal alpha peak it is worth wile to check, whether in this regime
the VCA gives similar results as the CPA. In Fig. \ref{fig5}
we show calculations with two Gaussians in CPA (Eq. (\ref{cpa3}) )
and VCA (Eq. (\ref{vca}) ) for different temperatures and a fixed
small width of the low-energy Gaussian. It can be seen that at high
frequency the beta peak is reproduced in the VCA, which demonstrates
that in the regime much above the alpha peak the VCA indeed agrees
to the CPA. It can be shown that in VCA the temperature and/or
frequency dependence reflexts the underlying barrier distribution
$P(E)$ with $E\propto -T\ln\omega$. So the peak at high omega
reflects the low-energy peak of $P(E)$. On the other hand, the
alpha peak - as discussed above - differs apprecially from the
principal maximum of $P(E)$.

\section{Conclusion}

In conclusion we can state that we have established a theory, which
combines a theory for the DC viscosity, low-temperature $\alpha$ and
$\beta$ relaxation and the high-frequency vibrational anomalies
within a unified framework. This has been achieved by assuming
that the viscous and elastic coefficients of Maxwell's theory
of viscoelasticity fluctuate in space according to a frozen
distribution of activation barriers. We have found an explanation
of the discrepance of the activation energies for diffusion and
viscosity in terms of the 
different percolative properties of the two heterogeneous
transport problems and a theory for the joint alpha and beta relaxation
below the glass transition.a

\section*{Acknowledgement}
W. S. is grateful for helpful discussions with
U. Buchenau,
J. C. Dyre,
W. G\"otze, 
A. Loidel,
T. Lunkenheimer, and
R. Schilling

\end{document}